\documentclass[twocolumn,aps,nofootinbib]{revtex4}
\usepackage{amsfonts, amssymb}
\usepackage{graphicx, epsfig,bm}
\usepackage{color}

\newcommand{\be}{\begin{equation}}
\newcommand{\ee}{\end{equation}}
\newcommand{\bea}{\begin{eqnarray}}
\newcommand{\eea}{\end{eqnarray}}

\newcommand{\pp}{.}
\newcommand{\vv}{,}

\def\fun#1#2{\lower3.6pt\vbox{\baselineskip0pt\lineskip.9pt
        \ialign{$\mathsurround=0pt#1\hfill##\hfil$\crcr#2\crcr\sim\crcr}}}

\newcommand\lsim{\mathrel{\rlap{\lower4pt\hbox{\hskip1pt$\sim$}}
    \raise1pt\hbox{$<$}}}
\newcommand\gsim{\mathrel{\rlap{\lower4pt\hbox{\hskip1pt$\sim$}}
    \raise1pt\hbox{$>$}}}

\def\dslash{\not{\hbox{\kern-2pt $\partial$}}}
\def\Dslash{\not{\hbox{\kern-4pt $D$}}}
\def\Oslash{\not{\hbox{\kern-4pt $O$}}}
\def\Qslash{\not{\hbox{\kern-4pt $Q$}}}
\def\pslash{\not{\hbox{\kern-2.3pt $p$}}}
\def\kslash{\not{\hbox{\kern-2.3pt $k$}}}
\def\qslash{\not{\hbox{\kern-2.3pt $q$}}}

 \newtoks\slashfraction
 \slashfraction={.13}
 \def\slash#1{\setbox0\hbox{$ #1 $}
 \setbox0\hbox to \the\slashfraction\wd0{\hss \box0}/\box0 }

\def\ee{\end{equation}}
\def\be{\begin{equation}}

\begin{document}
\setlength{\unitlength}{1mm}
\title{Constraints on neutrino -- dark matter interactions from cosmic microwave background and large scale structure data}

\author{Paolo Serra$^1$, Federico Zalamea$^{1,2}$, Asantha Cooray$^1$, Gianpiero Mangano$^3$, and Alessandro Melchiorri$^4$}

\affiliation{$^1$Center for Cosmology, Department of Physics and Astronomy,
 University of California, Irvine, CA 92697
CA 92697}

\affiliation{$^2$ Paris Diderot Physics Department, 10, rue Alice Domon et Leonie Duquet, 75205 Paris Cedex 13}

\affiliation{$^3$ Istituto Nazionale di Fisica Nucleare - Sezione di Napoli,
Complesso Universitario di Monte S. Angelo, I-80126 Napoli, Italy}

\affiliation{$^4$Physics Department and INFN, University of
Rome ``La Sapienza'', P.le Aldo Moro 2, 00185 Rome, Italy}


\begin{abstract}
We update a previous investigation of cosmological effects of a non-standard interaction between neutrinos and dark matter. Parameterizing the elastic-scattering cross section between the two species as a function of the temperature of the universe, the resulting neutrino-dark matter fluid has a non-zero pressure, which determines diffusion-damped oscillations in the matter power spectrum similar to the acoustic oscillations generated by the photon-baryon fluid. Using cosmic microwave background data in combination with large scale structure experiment results, we then put constraints on the fraction of the interacting dark matter component as well as on the corresponding opacity.
\end{abstract}
\bigskip

\maketitle

\section{Introduction}
Both the existence and the energy content of dark matter in the universe have been firmly estabilished by several observations (see \cite{Bertone2004} for a review), yet, after decades of research its nature is still unknown. The favored candidate for dark matter has three main feature namely, it seems to be a long-lived, cold and collisionless species. However, it is possible to relax some of these assumptions and in fact, several different dark matter candidates have been proposed through years (\cite{Jungman:1995df,axion}) and it is not even clear if only one species is responsible for the whole dark matter in the universe.\\
An interacting component has been proposed by many authors; couplings with a light scalar or pseudoscalar boson, as in the Majoron model has been investigated in \cite{hannestad1,hannestad2,pierpaoli,Dodelson2004} and interaction with neutrinos and more generally with the electromagnetic plasma have been considered for example, in \cite{boehm2002, boehm2003a, boehm2003b, Hooper2004, boehm2003c}.\\
In this paper we study the scenario where at least a fraction of dark matter is coupled to neutrinos. We parameterize the neutrino-dark matter interaction in terms of the opacity $P\equiv\langle\sigma_{dm-\nu}|v|\rangle/m_{dm}$, the ratio of the thermal averaged dark-matter-neutrino scattering cross section to the mass of the dark matter particle, where $P=Q \, a^{-2}$ with $Q$ a constant and $a$ the scale factor. We will briefly review the underlying theoretical basis of this choice in the next Section.\\  
The main observable consequence of such a coupling is a pattern of oscillations in the matter power spectrum, which can in principle mimic the baryon acoustic oscillations observed in the large scale clustering of galaxies \cite{Eisenstein2005,Percival2009}. The cosmic microwave background (CMB) radiation is  also affected by this interaction. Indeed, for sufficiently small scales, the temporal growth of a dark matter perturbation is lowered because of the coupling with neutrinos, and this determines a small enhancement of the small scale peaks  at a level which depends on both the strength of the interaction and the percentage of interacting dark matter. This effect can be also described in terms of a lowering of the neutrino "viscosity parameter" $c^2_{vis}$ \cite{Hu1998} from its standard value $c^2_{vis}=1/3$.\\
\begin{figure}[t]
\begin{center}
 \begin{tabular}{ccc}
\resizebox{80mm}{!}{\includegraphics{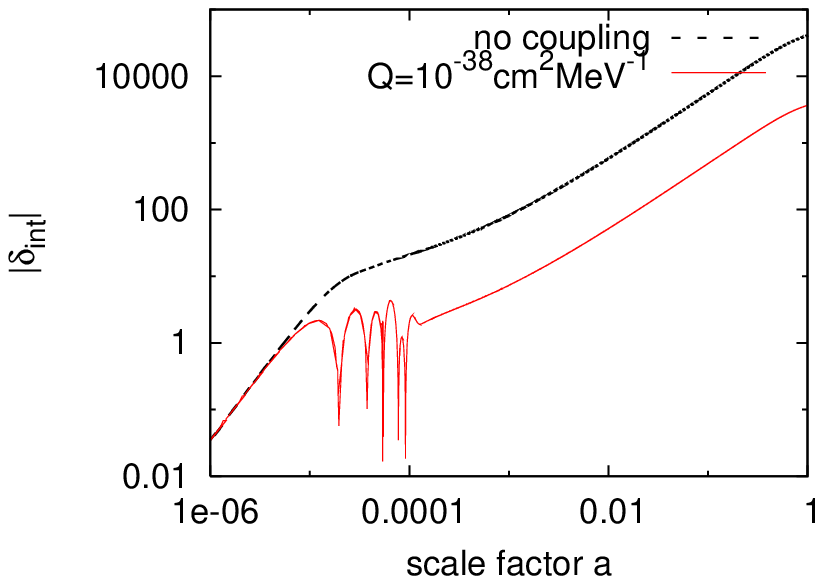}}\\
\resizebox{80mm}{!}{\includegraphics{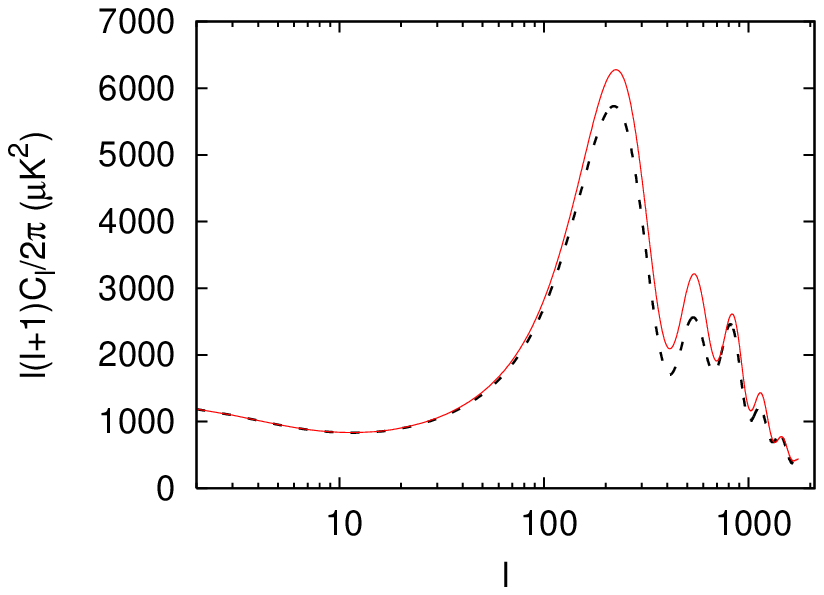}}\\
\resizebox{80mm}{!}{\includegraphics{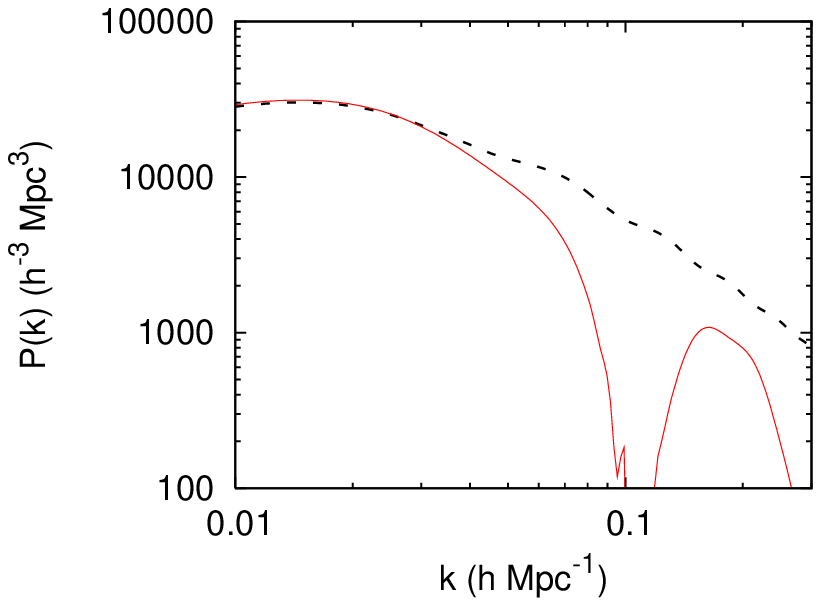}}\\
\end{tabular}
\caption{A perturbation of interacting dark matter (red line) of wavenumber $k=0.81\,h\,$Mpc$^{-1}$ is plotted against a perturbation of non-interacting dark matter (black dotted line) in the upper panel. The effects of this interaction are clearly seen in the angular power spectum of the CMB (middle panel) and on the matter power spectum (lower panel). We have chosen a very large value of Q to magnify the effect on the main cosmological observables.}
\end{center}
\end{figure}
This paper updates and improves the results of a previous work \cite{Mangano2006}. In fact, we present the results of a likelihood analysis based on a Monte Carlo Markov Chain sampling of the full cosmological parameter space, which allows us to take into account for possible degeneracies with the new parameters introduced in the analysis. To this aim, we use the latest CMB datasets available from the WMAP experiment \cite{Komatsu2009} in combination with large scale structure data from SDSS. 

The paper is organized as follows. In Section II we briefly discuss possible theoretical models for the dark matter-neutrino interaction and corresponding opacity due to scatterings. We refer to \cite{Mangano2006} for a more detailed discussion of these topics. In Section III we explain the method considered in the analysis and the data used to constrain both the strength of the interaction and the fraction of interacting dark matter. Finally we give our results and conclusions.       
\section{Neutrino-dark-matter scatterings}
Neutrino -- dark matter interactions can be modelled in several ways, depending on the spin of the dark matter candidates as well the specific coupling lagrangian. This issue is covered in details in \cite{boehm2003a,Mangano2006} to which we refer for more details, while we summarize in this Section the main general features of such interactions and report the thermal averaged cross section times velocity as function of the scale factor. The latter, denoted by $a$ is normalized to unity today, and scales as the inverse neutrino temperature. 

As was shown in \cite{Mangano2006}, if scatterings are supposed to affect the structure formation on large scales, it is unlikely that the relic abundance of these non standard dark matter candidates formed via the well known freezing phenomenon, as the pair annihilations would be quite efficient till recent times in depleting their number density, unless they represent a too light species. On the other hand, a different scenario has been also considered in the literature, where dark matter relic density today is the remnant of an initial particle -- antiparticle asymmetry produced in the early universe, thus similar to the mechanism of baryogenesis which leads to a baryon density today much larger than what is expected by freezing of strong interactions alone \cite{kaplan,Hooper:2004dc}. This possibility may also account for the intriguing similarity between the values of $\Omega_b$ and $\Omega_{dm}$ we observe today, which differ by a factor five only, a feature which may call for a common mechanism for their formation. In the following we therefore, consider the case of a non self--conjugated fermion or scalar particle $\psi$ with a conserved global $U(1)$ charge, at least in the low energy scale regime, say below their mass scale, which also corresponds to the relevant stages for structure formation we are interested in. 

In case of a spin zero species, the interaction lagrangian  can be chosen as a Yukawa term
\be {\cal L}_{int} = h \overline{F}_R \nu_L \psi + h.c. \vv \label{scalar1} \ee
with $F$ a spinor field, or via coupling with an intermediate vector-boson field $U_\mu$
\bea  {\cal L}_{int} &=& i g (\psi^* \partial^\mu \psi - \psi \partial^\mu \psi^*) U_\mu + g^2 \psi^* \psi U_\mu U^\mu \nonumber \\
&+& g_\nu \overline{\nu} _L\gamma^\mu  \nu_L U_\mu \pp \label{scalar2} \eea
In both cases $F$ and $U$ fields will be assumed to have mass larger than $\psi$, to prevent fast $\psi$ decay at tree level, see \cite{Mangano2006}. 

In case of spin 1/2 dark matter, one has
\be
 {\cal L}_{int} = h \overline{\psi}_R \nu_L F + h.c. \vv \label{fermion1} \ee
 with $F$ a scalar field or finally,
 \bea
 {\cal L}_{int} &=& g (c_L \overline{\psi}_L \gamma^\mu \psi_L + c_R \overline{\psi}_R \gamma^\mu \psi_R) U_\mu \nonumber \\
 &+& g_\nu \overline{\nu}_L \gamma^\mu \nu_L U_\mu \pp \label{fermion2} \eea
 We notice that, as we will see in the next Section, bounds on dark matter -- neutrino scattering cross section from CMB and large scale data, correspond to a mass scale for $\psi$ as well as the other field involved in the interaction lagrangian ($F$ or $U$) larger than MeV. Thus, we will assume this lower bound in the following, $m_{dm} \geq$ MeV, $m_{F,U} \geq$ MeV.

 It is quite easy to compute the thermal averaged scattering cross section corresponding to these interaction terms in the non-relativistic limit for the dark matter particle $\psi$ and $F$ or $U$. For example in the case of Eq. (\ref{scalar1}) or (\ref{fermion1}) and (\ref{scalar2}) one gets, respectively
 \bea
  \langle \sigma_{dm-\nu} |v| \rangle &\sim & |h|^4 \frac{T_\nu^2}{(m_F^2-m_{dm}^2)^2} \vv \\
   \langle \sigma_{dm-\nu} |v| \rangle &\sim & g^2 g_\nu^2 \frac{T_\nu^2}{m_U^4} \pp
   \eea
 The first result holds for $m_F \neq m_{dm}$, otherwise cross section takes a constant value, as for the well known Thomson cross section. This special case has been considered in details in \cite{Mangano2006}, and will not be worked out here,  as it is quite fine-tuned. 
 Similar results hold for a spinorial dark matter interacting via an intermediate gauge boson $U$
 \be
  \langle \sigma_{dm-\nu} |v| \rangle \sim  g^2 g_\nu^2 \left( c_L^2+c_R^2-c_L c_R \right) \frac{T_\nu^2}{m_U^4} \vv 
\ee
  We see that for both Yukawa-type or intermediate gauge boson interaction, the opacity assumes a simple expression, of the form
 \be  P  \sim \frac{1}{M^4 m_{dm}} T_\nu^2 = \frac{T_\nu^{(0)2}}{M^4 m_{dm}} a^{-2} \equiv Q \,\, a^{-2} \vv
 \ee
 with $M$ a mass scale and $Q$ a constant parameterizing the strength of interaction
 \be 
 Q = \frac{1}{(M/ \mbox{MeV})^4} \frac{1}{m_{dm}/ \mbox{MeV}} \cdot 10^{-41}\,\,\,  \mbox{cm}^2 \mbox{MeV}^{-1} \vv
 \ee
 where we have used the value of neutrino temperature today $T_\nu^{(0)}$.
 \begin{table}
\caption{Mean values and marginalized confidence levels for the
cosmological parameters of interest for WMAP+SDSS. When not otherwise stated, all constraints are at $68\%$ c.l..}
\begin{center}
\begin{tabular}{|c|c|}
\hline
Parameter & WMAP+SDSS  \\
\hline
$\Omega_bh^2$&$0.02271\pm0.00062$\\
$\Omega_{\rm c}h^2$&$0.059_{-0.019}^{+0.020}$ \\
$\Omega_{\rm x}h^2$&$<0.093 (2\sigma)$ \\
$\Omega_{\Lambda}$&$0.754\pm 0.018$\\
$n_s$&$0.964\pm0.014$\\
$\tau$&$0.089\pm0.017$\\
$\Delta^2_R$&$(2.38\pm0.10)\cdot10^{-9}$\\
\hline
$H_0$&$72.7\pm1.8$\\
$z_{reion}$&$10.4\pm1.3$\\
$t_0$&$13.69\pm0.13$\\
$A_{SZ}$&$1.08_{-0.51}^{+0.92}$\\
\hline

\end{tabular}
 \label{table:1}
 \end{center}
 \end{table}

\begin{figure}[!t]
\begin{center}
\includegraphics[scale=0.5]{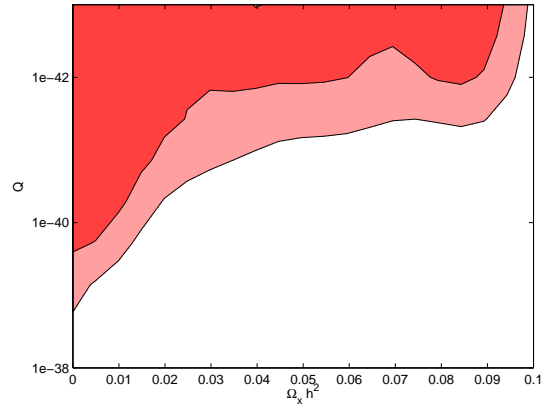}
 \caption{Constraints in the  cross section parameter $Q$ versus $\Omega_X h^2$ plane from the combination of 
WMAP and SDSS data. Large values for the neutrino-dark matter interaction are always compatible with the standard picure, provided that the energy content of the interacting dark matter is small enough.}
\label{fig.4}
\end{center}
\end{figure}

\section{Analysis}
In the early universe, the opacity $P$ is strong enough to determine acoustic oscillations in the coupled neutrino-dark matter fluid, just like the usual electromagnetic force does with the photon-baryon fluid before recombination. In order to include a neutrino-dark matter interaction, we modify the standard Eulero equation for dark matter, resulting in an analogue of that for baryon matter
\begin{equation}
\label{cdm2} \dot{\theta}_{dm} = -{\dot{a}\over a}\theta_{dm} +
{4\rho_\nu \over 3\rho_{dm}}a \, n_{dm}\,\sigma_{dm-\nu}
(\theta_\nu-\theta_{dm}) \vv \\
\end{equation}
where momentum conservation in scattering processes has been accounted for. With $\theta$ we denote the velocity perturbations and the product $a\,n_{dm}\,\sigma_{dm-\nu}$ quantifies the scattering rate of neutrinos by dark matter. As in the case of the baryon-photon interaction, we can neglect both the shear contribution $k^{2}\sigma$ and the term $c_{dm}^{2}k^{2}\delta$, where $c_{dm}$ is the sound speed of the dark-matter fluid. The hierarchy of equations for neutrinos has been modified accordingly.\\
As we can see from the upper panel of Fig.~1, any interacting dark matter perturbation of sufficiently small scale will be affected by the coupling with neutrinos. When it enters the horizon, the mode will experience a series of acoustic oscillations while coupled with neutrinos, rather than growing sligthly in radiation dominated epoch and then more rapidly in matter dominated epoch. This determines a pattern of oscillations in the matter power spectrum (lower panel) as well as an enhancement of the peaks in the angular power spectrum (middle panel).\\
We make use of the publicly available Markov Chain Monte Carlo (MCMC) package CosmoMC \cite{Lewis2002} with a convergence diagnostic based on the Gelman and Rubin statistic \cite{Gelman1992} to put constraints on the cosmological parameters. We consider a flat $\Lambda$CDM model described by the following set of nine cosmological parameters:
\begin{equation}
 \label{parameter}
      \{\omega_{b}, \omega_{c},\omega_{X},
      \Theta_{s}, \tau,  n_{s}, \log[10^{10}A_{s}],Q, A_{SZ} \}~,
\end{equation}
where $\omega_{b}$ ($\equiv\Omega_{b}h^{2}$), $\omega_{c}$ ($\equiv\Omega_{c}h^{2}$), $\omega_{X}$ ($\equiv\Omega_{X}h^{2}$) are the baryon, standard collisionless cold dark matter and interacting cold dark matter densities relative to the critical density respectively, $\Theta_{s}$ is the ratio of the sound horizon to the angular
diameter distance at decoupling, $\tau$ the optical depth to reionization and $A_{s}$ and $n_s$ are the amplitude of the primordial spectrum and the spectral index. They are both evaluated at $k=0.002$ Mpc$^{-1}$. The parameter $Q$ (measured in units of cm$^2$ MeV$^{-1}$) quantifies the strength of the neutrino--dark matter interaction, as discussed in the previous section. Finally, we take into account for the contribution to the power spectrum by Sunyaev-Zel'dovich fluctuations \cite{Sunyaev1970} by adding the predicted template spectrum from \cite{Komatsu2002}, multiplied by the constant amplitude $A_{SZ}$.
We use data and likelihood code from the WMAP team's 5 year release \cite{Komatsu2009,Dunkley2009} (both temperature and polarization) in combination with the SDSS LRG data release \cite{Tegmark2006}.\\
In Fig.~2 we plot the combined constraints on both the parameter $Q$ and the energy content in interacting dark matter. As expected, there is a clear correlation between these two parameters as the lower the energy content in interacting dark matter the weaker the constraint on the strength of the interaction. It has been shown in \cite{Mangano2006} that the baryon acoustic oscillations due to the coupled photon-baryon plasma and seen in the large scale clustering of matter \cite{Eisenstein2005,Percival2009} can mimic the coupled neutrino-dark matter fluid in a universe with a lower energy content in baryon, seriously affecting constraints on $\Omega_bh^2$ derived by large scale structures data only. We find here that  when a full parameter space analysis is performed using additionaly data from the CMB, we always recover the standard constraints for the baryon density. Of course the energy content in standard dark matter is smaller than the usual amount simply because we are splitting all the dark matter in an interacting and a non interacting component. All the other values for the cosmological parameters are compatible with the standard values at $1\sigma$.


\section{Conclusions}
In this paper we have investigated the effects of a non-standard interaction between neutrinos and dark matter using the to date available data from WMAP and SDSS Collaborations on cosmic microwave background and large scale structures. These effects are the presence of diffusion -- damped oscillations in the matter power spectrum and a small enhancement of the peaks in the cosmic microwave background angular power spectrum. Parameterizing the opacity parameter as $Q \,a^{-2}$ with $Q$ a constant, which holds in a large class of models for neutrino -- dark matter interactions, we found that  the bound on $Q$ becomes stronger as the percentage $\Omega_X h^2$ of interacting dark matter component grows. Typical upper bound on $Q$ is of the order of $10^{-42} \div 10^{-40}$ cm$^2$ MeV$^{-1}$, while the bound becomes weaker for very low values of $\Omega_X$; we note that cosmological constraints on neutrino masses can be affected because of possible degeneracies.

\smallskip
This research was funded by NSF CAREER AST-0605427 and by LANL IGPP-08-505.

\end{document}